\documentclass{aa}
\usepackage{graphicx}
\usepackage{natbib}
\usepackage{color}
\usepackage{float}
\usepackage{txfonts}
\usepackage{pdflscape}

\newcommand{\mlc}{\multicolumn{1}{c}}

\begin{document}

\title{Hunt for extremely eccentric eclipsing binaries}

\author{Zasche,~P.~\inst{1},
        Henzl,~Z.~\inst{2},
        Ma\v{s}ek,~M.~\inst{2,3}
        }

\offprints{Petr Zasche, \email{zasche@sirrah.troja.mff.cuni.cz}}

 \institute{
  $^{1}$ Astronomical Institute, Charles University, Faculty of Mathematics and Physics, V~Hole\v{s}ovi\v{c}k\'ach 2, CZ-180~00, Praha 8, Czech Republic\\
  $^{2}$ Variable Star and Exoplanet Section of the Czech Astronomical Society, Vset\'{\i}nsk\'a 941/78, CZ-757 01 Vala\v{s}sk\'e Mezi\v{r}\'{\i}\v{c}\'{\i}, Czech Republic \\
  $^{3}$ FZU - Institute of Physics of the Czech Academy of Sciences, Na Slovance 1999/2, CZ-182~21, Praha, Czech Republic
 }

\titlerunning{Extremely eccentric eclipsing binaries}
\authorrunning{Zasche et al.}

  \date{Received \today; accepted ???}

\abstract{We report the very first analysis of 27 eclipsing binary systems with high eccentricities
that sometimes reach up to 0.8. The orbital periods for these systems range from 1.4 to 37 days, and
the median of the sample is 10.3 days. Star CzeV3392 (= UCAC4 623 022784), for example, currently is
the eclipsing system with the highest eccentricity (e=0.22) of stars with a period shorter than 1.5
days. We analysed the light curves of all 27 systems and obtained the physical parameters of both
components, such as relative radii, inclinations, or relative luminosities. The most important
parameters appear to be the derived periods and eccentricities. They allow constructing the
period-eccentricity diagram. This eccentricity distribution is used to study the tidal circularisation
theories. Many systems have detected third-light contributions, which means that the Kozai-Lidov cycles
might also be responsible for the high eccentricities in some of the binaries. }


\maketitle

\section{Introduction} \label{intro}

Eclipsing binaries (hereafter EBs) still play a crucial role in astrophysics today
\citep{2012ocpd.conf...51S}, mainly because they are used as calibrators in stellar evolutionary models
\citep{2010A&ARv..18...67T}. However, another use of EBs is also widely found in the literature. In
addition to their use as distance indicators, they can also serve as ideal sources for statistical
studies of stellar populations, such as studies of period distributions in various stellar populations,
mass ratio distributions, or period-eccentricity relations (see e.g. \citealt{2008MNRAS.389..925T}).

Period-eccentricity relations are studied mainly because a certain limit of the maximum value of
eccentricity for a particular period is expected. The circularisation is a very slow process, but it is
also inevitable. It causes all eccentric orbits to be more circular. The efficiency of this effect
strongly depends on the relative sizes of the stars, that is, their $R/a$ ratios. This means that the
systems that are close enough (compact systems) should be well circularised or are still very young. On
the other hand, longer-period systems should still have some non-zero eccentricity. All of these were
studied several times in the past (using different binary samples) as a consequence of the tidal
interaction in these systems, see for example \cite{2019EAS....82..127S} or \cite{2008EAS....29....1M}.
For the eclipsing binaries, the Kepler satellite and the studies of eclipsing binaries detected in its
fields were very important. Quite recently, a study \citep{2021A&A...648A.113B} analysed the red giants
in eclipsing binaries and discussed the circularisation theories in these evolved stars. Another paper
\citep{2016ApJ...824...15V} studied the sample of Kepler binaries with respect to their temperatures
and found an agreement with the theory that hotter stars have longer circularisation times.

Therefore we decided to focus our attention on the still rather sparsely populated areas in the
period-eccentricity diagram near the higher limit of eccentricity for the particular period. These new
data can bring us some new limits for the maximum eccentricity and the tidal circularisation theories,
or possibly for the study of Kozai-Lidov cycles. Such effect is able to excite the eccentricity to
higher values, or even break up the whole system. A necessary condition is a third component in the
system in addition to the inner eclipsing pair, however. A weak indication of such a putative third
body should also be gained through an analysis of the light curves (hereafter LCs) when a conclusive
level of third-light contribution is detected.

\section{Current status of the topic}

Only a rather limited number of papers on the P-e diagram for eclipsing binaries is published so far,
with only a few attempts to identify these higher eccentric systems. For example the well-known
catalogues of eccentric eclipsing binaries published in 2007 by \citeauthor{2007MNRAS.378..179B} that
were later updated by \cite{2018ApJS..235...41K} are surprisingly very incomplete in this aspect. The
authors studied only the $O-C$ diagrams and apsidal motions, but the light curve solutions published
for some of the systems were not included, especially those with higher eccentricities. In particular,
the apsidal motion and the orbital period are related (the rapid apsidal motion is usually present in
shorter-period systems), therefore these systems with high eccentricities were missed because they are
not so interesting for observers. Moreover, \cite{2018ApJS..235...41K} were a rather dubious source of
information on eccentricity at times. For example, the system with the highest eccentricity in
\cite{2018ApJS..235...41K} is V345 Lac with e=0.54, but this information was later revised to a much
lower value of 0.26 by \cite{2019AcA....69...63W}. This is a quite typical situation when the method of
the $O-C$ diagram analysis alone is used to derive the eccentricity instead of a more reliable light
curve analysis. The most eccentric system in \cite{2018ApJS..235...41K} therefore is the star V1143 Cyg
(= HD 185912, see \citealt{2019AJ....158..218L}) with a period of 7.64~days and an eccentricity of
0.538.

The P-e relation was studied several times for spectroscopic binaries (see e.g.
\citealt{2002AJ....124.1144L}, or \citealt{2005ApJ...620..970M}) or other types of systems such as
exoplanets \citep{2008ApJ...686..603J}. However, because it is relatively easy and straightforward to
derive the eccentricity in case of eclipsing binaries, we decided to use these objects. The eclipsing
systems were used to detect of higher eccenricities, for instance using the ASAS
\citep{2004AcA....54..153P} photometric surveys by \cite{2014MNRAS.441..343S}, which appears to be one
of only a few dedicated studies focusing on the detection of high eccentricities. The authors found the
highest eccentricity for system ASAS 144242-5904.1 with e=0.64, which was the maximum eccentricity
found for eclipsing binary at that time. The rather questionable eclipsing system IO Com with its
eccentric orbit e=0.69 and P=53~d is still quite problematic because no secondary eclipse has been
detected \citep{2014BaltA..23...27T}. The rich and precise data provided by the Kepler satellite
\citep{2010Sci...327..977B} were analysed later and yielded a large number of detached and highly
eccentric systems, one of which has an eccentricity of 0.845, but an orbital period of 265~d
\citep{2017AJ....154..105K}. However, these highly eccentric systems usually have rather long orbital
periods, and the most eccentric system with an orbital period shorter than 40 days has an eccentricity
of 0.702. Unfortunately, these longer-period systems have only little effect on our study of the P-e
diagram because the most interesting curvature lies in periods shorter than 50 days. This was the main
motivation for our study.

\section{Selected systems}

We chose for our analysis binaries that satisfied the following criteria: i) They were never studied
before (to enlarge the sample of already known binaries with high eccentricities), ii) they have good
phase coverage of their light curves (only with well-covered LC are we able to derive the eccentricity
conclusively enough), and iii) their high eccentricities are obvious. The last point comes easily from
the fact that with the publicly available databases today, we can identify systems with high
eccentricities only when the photometric data are plotted with proper period. These most eccentric
systems show two signs of their eccentric orbits. First, the displacement of the primary and secondary
minima out of 0.0 and 0.5 phases, respectively. Second, the very different eclipse durations. A
combination of these two approaches applies for most of the systems we present here.

All binaries presented and analysed below were found using one of these two methods. First, six systems
were identified in the OGLE fields of the LMC galaxy. The whole set of 26121 eclipsing binaries
identified in the OGLE III dataset \citep{2011AcA....61..103G} was scanned and the most eccentric
systems were chosen. Then we used OGLE III and OGLE IV data \citep{2016AcA....66..421P} for the
subsequent analysis. A second part of our sample are the systems that were discovered by chance in the
monitored fields of some already known system by the authors M.M. and Z.H. For these new binaries the
new TESS photometry \citep{2015JATIS...1a4003R} was downloaded and analysed. For obtaining the
photometry from the raw TESS data, we used the tool {\tt{lightkurve}}  \citep{2018ascl.soft12013L}.
There was the usual problem with these TESS data that the pixels are too large and the photometry is
sometimes contaminated with close components (which usually has nothing to do with the eclipsing binary
itself). Sometimes the method of flattening these TESS fluxes provides rather problematic trends and
artefacts. Therefore we have to be very cautious when comparing the TESS photometric data with other
ground-based photometry, especially its eclipse depths. For these reasons, we plot the TESS photometry
only relatively (i.e. the out-of-eclipse magnitude was set to 0.0). For only one system were we not
able to obtain any TESS data, therefore we used for the LC analysis the other available photometry from
ASAS \citep{2002AcA....52..397P}, and ASAS-SN (\citealt{2014ApJ...788...48S}, and
\citealt{2017PASP..129j4502K}) databases.

For basic information about the selected stars, see Table \ref{systemsInfo}. Their identification in
various catalogues is presented. Here we would like to emphasize the steadily growing number of new
eclipsing systems (more than 2200 at present) discovered and included in the Czech Variable Star
Catalogue, CzeV \citep{2017OEJV..185....1S}. Moreover, the position on the sky and magnitude
out-of-eclipse is provided in the Table \ref{systemsInfo} together with some temperature or spectral
information, if available.

\begin{figure*} 
  \centering
  \includegraphics[width=0.845\textwidth]{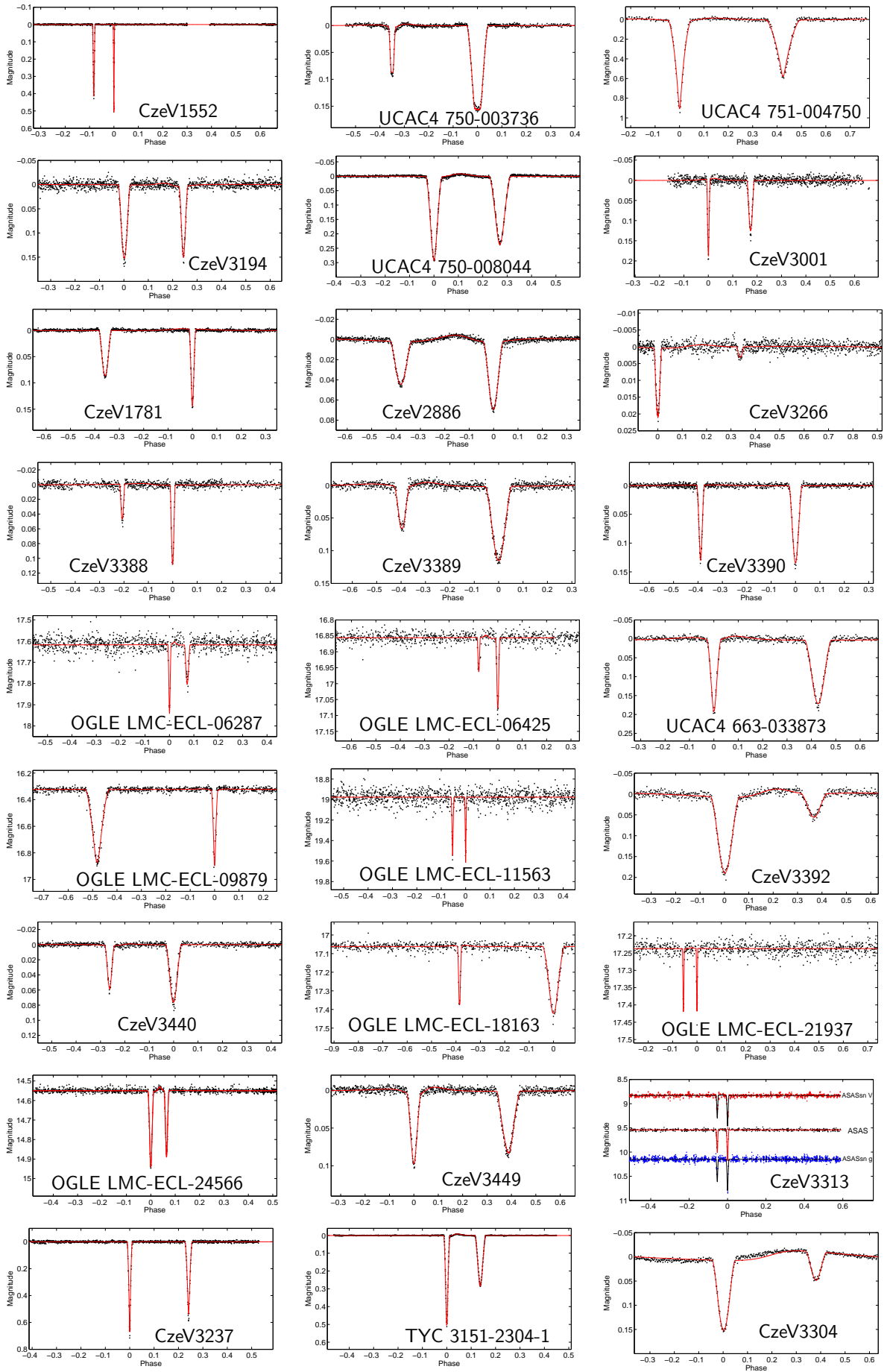}
  \caption{Final light curves from our analysis.}
  \label{FigLCs}
\end{figure*}

\begin{table*}
  \caption{Individual systems.}  \label{systemsInfo}
  \scalebox{0.81}{
  \begin{tabular}{c c c c c c}\\[-6mm]
\hline \hline
  Target name         &  Other name                 &RA [J2000.0]& DE [J2000.0]& Mag$_{max}$ $^*$ &  Spectral/Temperature information $^{**}$     \\
 \hline
 CzeV1552             & UCAC4 745-002478            & 00 19 18.9 & +58 54 56.0 &  13.19 (V)       & T$_{eff} = 5372$ K \citep{2018AaA...616A...1G} \\
 UCAC4 750-003736     & ASASSN-V J002153.45+595840.9& 00 21 53.5 & +59 58 40.9 &  13.61 (V)       & T$_{eff} = 6978$ K \citep{2019AJ....158...93B} \\
 UCAC4 751-004750     & ASASSN-V J002851.87+600443.5& 00 28 51.9 & +60 04 43.5 &  15.00 (V)       & T$_{eff} = 6678$ K \citep{2019AJ....158...93B} \\
 CzeV3194             & UCAC4 750-006133            & 00 39 06.7 & +59 51 29.2 &  15.73 (V)       & T$_{eff} = 4446$ K \citep{2018AaA...616A...1G} \\
 UCAC4 750-008044     & ASASSN-V J005130.07+595845.6& 00 51 30.1 & +59 58 45.6 &  12.89 (V)       & T$_{eff} = 5764$ K \citep{2018AaA...616A...1G} \\
 CzeV3001             & UCAC4 753-010281            & 00 51 38.5 & +60 28 19.4 &  14.78 (V)       & T$_{eff} = 5911$ K \citep{2018AaA...616A...1G} \\
 CzeV1781             & UCAC4 751-009245            & 00 53 17.8 & +60 09 41.7 &  12.86 (V)       & T$_{eff} = 5287$ K \citep{2018AaA...616A...1G} \\
 CzeV2886             & UCAC4 750-008580            & 00 53 48.2 & +59 51 34.6 &  12.79 (V)       & T$_{eff} = 8031$ K \citep{2019AJ....158...93B} \\
 CzeV3266             & UCAC4 765-011189            & 00 57 47.6 & +62 59 02.8 &  12.86 (V)       & T$_{eff} = 7978$ K \citep{2019AJ....158...93B} \\
 CzeV3388             & UCAC4 752-011930            & 01 00 12.2 & +60 12 21.6 &  13.67 (V)       & T$_{eff} = 7197$ K \citep{2019AJ....158...93B} \\
 CzeV3389             & UCAC4 755-012648            & 01 05 11.2 & +60 56 36.6 &  15.59 (V)       & T$_{eff} = 4981$ K \citep{2018AaA...616A...1G} \\
 CzeV3390             & UCAC4 753-014055            & 01 07 10.1 & +60 31 57.7 &  14.04 (V)       & T$_{eff} = 7751$ K \citep{2019AJ....158...93B} \\
 OGLE LMC-ECL-06287   & OGLE LMC-SC14 165496        & 05 03 52.2 & -68 59 18.4 &  17.62 (I)       & (B-V)$_0$ = -0.273 mag \citep{2002ApJS..141...81M} \\
 OGLE LMC-ECL-06425   & OGLE LMC116.5 198           & 05 04 12.0 & -67 17 27.4 &  16.86 (I)       & (B-V)$_0$ = -0.160 mag \citep{2004AJ....128.1606Z} \\
 UCAC4 663-033873     & ASASSN-V J050901.74+423234.0& 05 09 01.7 & +42 32 34.0 &  14.78 (V)       & T$_{eff} = 5869$ K \citep{2018AaA...616A...1G} \\
 OGLE LMC-ECL-09879   & OGLE LMC107.8 14712         & 05 12 20.1 & -67 07 31.7 &  16.32 (I)       & (B-V)$_0$ = -0.115 mag \citep{2004AJ....128.1606Z} \\
 OGLE LMC-ECL-11563   & OGLE LMC101.6 1597          & 05 16 33.2 & -68 37 50.2 &  18.98 (I)       & (B-V)$_0$ = -0.185 mag \citep{2004AJ....128.1606Z} \\
 CzeV3392             & UCAC4 623-022784            & 05 20 30.7 & +34 26 33.6 &  15.36 (V)       & T$_{eff} = 7312$ K \citep{2019AJ....158...93B} \\
 CzeV3440             & UCAC4 621-023131            & 05 23 14.8 & +34 00 53.7 &  14.43 (V)       & B4V \citep{2019ApJS..241...32L} \\
 OGLE LMC-ECL-18163   & OGLE LMC169.6 119123        & 05 31 23.4 & -69 44 59.2 &  17.06 (I)       & (B-V)$_0$ = -0.042 mag \citep{2004AJ....128.1606Z} \\
 OGLE LMC-ECL-21937   & OGLE LMC176.3 38133         & 05 39 57.8 & -69 44 44.7 &  17.24 (I)       & (B-V)$_0$ = -0.236 mag \citep{2004AJ....128.1606Z} \\
 OGLE LMC-ECL-24566   & MGSD LH 117 14              & 05 48 54.6 & -70 02 29.8 &  14.55 (I)       & O6.5 \citep{{1989AJ.....97..107M}} \\
 CzeV3449             & UCAC4 297-025583            & 07 35 14.4 & -30 36 06.4 &  14.98 (V)       & T$_{eff} =7590$ K \citep{2019AJ....158...93B} \\
 CzeV3313 = HD 61302  & UCAC4 371-033137            & 07 38 05.7 & -15 54 56.3 &  10.12 (V)       & A3/5II \citep{1988mcts.book.....H} \\
 CzeV3237             & UCAC4 665-087278            & 20 12 15.5 & +42 58 38.4 &  15.46 (V)       & T$_{eff} = 6328$ K \citep{2019AJ....158...93B} \\
 TYC 3151-2304-1      & UCAC4 644-089491            & 20 15 54.8 & +38 46 14.2 &  11.77 (V)       & OB \citep{1952ApJ...115..459N} \\
 CzeV3304             & UCAC4 755-077889            & 23 12 08.0 & +60 59 30.2 &  14.60 (V)       & T$_{eff} = 7045$ K \citep{2019AJ....158...93B} \\
 \hline
\end{tabular}}\\
 {\small Notes: $^*$ - Out-of-eclipse magnitude, $I_{mag}$ from the OGLE survey \citep{2011AcA....61..103G}, $V_{mag}$ from the UCAC4 catalogue \citep{2013AJ....145...44Z}, or the Guide Star Catalog II \citep{2008AJ....136..735L}, $^{**}$ - Spectral estimate or dereddened photometric indices from \cite{2002ApJS..141...81M}, and \cite{2004AJ....128.1606Z}.}
\end{table*}

\section{Analysis}

For the whole analysis we used the software {\sc PHOEBE}, ver 0.32svn \citep{2005ApJ...628..426P}. It
uses a classical Roche-based model by \cite{1971ApJ...166..605W}, with its later modifications. The
eccentric orbit was also implemented. The only unknown issue therefore remains for the start of
analysis: the period. However, the periodicity of these selected stars was easily detectable from the
photometry. The TESS data provide a superb coverage and precision, while the OGLE database even
provides a period directly for every detected eclipsing binary.

For the analysis we applied several simplifications. Due to missing spectroscopy and only very limited
information about these stars, we applied the following assumptions: i) the mass ratio was set to
unity, ii) the synchronicity parameters were also set to pseudo-synchronous rotation, and iii) albedo
and gravity brightening coefficients were set to their suggested values according to the particular
temperature of the stars. The most problematic issue remains the effective temperature of the primary
component, which has to be set and remain fixed during the whole fitting process. Because the
spectroscopy is not available, the only piece of information we have are the photometric indices, which
can be used for a rough estimate of the temperatures. For the stars in our Galaxy using the TESS data,
we usually used the temperature estimate based on the GAIA DR2 data \citep{2018AaA...616A...1G}, while
for the stars in LMC using the OGLE data, we tried to estimate roughly the effective temperatures based
on the dereddened photometric indices as published by \cite{2004AJ....128.1606Z} and
\cite{2002ApJS..141...81M}. We are aware that our main goal is to derive the period and eccentricity,
and not to derive the precise physical parameters of the components. Therefore these simplifications
are probably not crucial.

We did not include any spots for modelling either. The only system showing some additional asymmetry is
CzeV3304. Its LC cannot be properly modelled with eccentricity and reflection (albedo and
gravity-brightening effects).

\section{Results}

All the results of the systems we analysed are plotted together in Fig. \ref{FigLCs}, where all the
final fits are given together with the data. The widths of the eclipses are rather narrow at times, but
it is important to plot the whole phase of the light curves, not only a part of it zoomed near the
min-eclipses. Because very narrow eclipses cover only a small fraction of the whole period, the proper
primary and secondary eclipses are sometimes covered with only a few data points. This is still
adequate for a precise derivation of the eccentricity value for the particular system, however. The
results of the fittings are given in Table \ref{LCparam} together with some basic parameters such as
relative radii $R/a$ and also the periods and eccentricities of all systems. The higher values of the
eccentricities for the longer-period binaries is clearly visible. See the more detailed discussion of
the P-e relation in the next section.

We plotted all light curves shifted in phase to always have the deeper minimum in 0.0, while the
primary and secondary eclipses are also given in Table \ref{LCparam}. These ephemerides for the primary
and secondary should be used by future observers to plan their observations of both eclipses
because the apsidal motion is negligible here. The hotter component is always the primary component, that is,
the component with the derived fixed temperature from the values given in Table \ref{systemsInfo}.

Although the individual masses are only roughly estimated and the mass ratio can only barely be derived
for these detached systems \citep{2005Ap&SS.296..221T}, our fits are thought to be reliable. They
should be taken as good starting points for some future more detailed analyses of these systems,
especially when good spectroscopy is available (then the need of a mass ratio fixed to unity would
become obsolete). However, the missing mass ratio information can led to rather problematic results
(e.g. the more luminous component is hotter, but smaller) that might also affect the resulting $p$-$e$
values. The question is whether a problematic assumption in the beginning can lead to incorrect
eccentricity values and if some significant offset of $e$ can be caused by our method. To determine
this, we tried the following test. A far more realistic LC solution (following the assumption of the
main-sequence components) can be obtained using the method of mass ratio derivation by
\cite{2003MNRAS.342.1334G}. This method uses the inferred luminosities of the two components to
calculate the mass ratio, which is then more realistic. We used this method for the first two systems
in our sample. For the first system, it resulted in exactly the same numbers as our simplified approach
(as expected because the two components are very similar to each other). For the second system, UCAC4
750-003736, however, this more sophisticated approach led to $q=0.55$ and $e=0.57$. This eccentricity
is even higher than the original ($e=0.54$), even outside the original error bars. This means that the
original uncertainties of the eccentricities are probably rather underestimated, a typical situation
when using {\sc PHOEBE}.

The detection of the third light (at a non-negligible level) for most of the systems was quite
remarkable. About two-thirds of all binaries here appear to contain some additional component. This is
a rather high fraction. Not all of them should be considered as triples in nature (because the quite
large TESS pixels also cause close stars in the field to contribute to the signal, and a similar
situation is known for the OGLE superdense LMC fields), but this result probably means that for a
significant fraction of the stars, some triple-star dynamics cannot be completely excluded.

\begin{table*}
 \caption{Parameters of the light-curve fits.}
 \label{LCparam}
 \footnotesize
 \centering \scalebox{0.67}{
 \begin{tabular}{l c c c c c c c c c c c c c}
 \hline\hline
  System             &    $i$       &  \mlc{$T_1$}  & \mlc{$T_2$} &   $L_1$    &   $L_2$    &   $L_3$   &  $R_1/a$  &  $R_2/a$  &  $T_{prim}$ &  $T_{sec}$  & \mlc{$P$ [d]}&   $e$     & $\omega$ \\
                     &    [deg]     &  \mlc{[K]}    & \mlc{ [K]}  &   [\%]     &    [\%]    &   [\% ]   &           &           & [JD-2450000]& [JD-2450000]&              &           & [deg]    \\
 \hline
 CzeV1552            & 88.83 (0.13) &  5400 (fixed) &  5247 (115) & 53.2 (0.6) & 46.8 (0.6) & 0         & 0.019 (1) & 0.019 (1) &8778.464 (2) &8775.616 (2) & 34.5230 (3)  & 0.754 (3) &203.2 (0.1) \\
 UCAC4 750-003736    & 89.98 (0.20) &  6978 (fixed) &  5778 (209) & 42.2 (0.6) &  7.0 (0.7) & 50.8 (0.7)& 0.105 (3) & 0.058 (2) &8953.102 (1) &8949.499 (1) & 10.29037 (2) & 0.539 (5) &292.5 (0.2) \\
 UCAC4 751-004750    & 89.85 (0.34) &  6678 (fixed) &  6239 (212) & 56.2 (2.0) & 43.8 (2.1) & 0         & 0.174 (3) & 0.185 (4) &8611.641 (2) &8612.527 (2) &  2.08790 (5) & 0.233 (9) &120.1 (0.4) \\
 CzeV3194            & 83.06 (0.14) &  4446 (fixed) &  3446 (304) & 67.0 (3.2) & 28.0 (2.7) & 5.0 (1.4) & 0.086 (4) & 0.094 (3) &8493.567 (1) &8494.659 (1) &  4.4699 (1)  & 0.478 (6) &213.4 (0.5) \\
 UCAC4 750-008044    & 85.33 (0.12) &  5764 (fixed) &  5690 (92)  & 26.5 (0.5) & 39.3 (0.6) & 34.2 (0.9)& 0.114 (2) & 0.142 (2) &8956.195 (1) &8957.184 (1) &  3.64441 (4) & 0.393 (3) &157.4 (0.3) \\
 CzeV3001            & 86.28 (0.11) &  6000 (fixed) &  4564 (216) & 73.5 (1.7) & 19.4 (3.3) &  7.1 (4.7)& 0.051 (3) & 0.044 (2) &8947.329 (2) &8949.640 (2) & 13.2511 (1)  & 0.677 (2) &314.0 (0.2) \\
 CzeV1781            & 85.38 (0.15) &  5287 (fixed) &  4421 (85)  & 57.2 (0.4) & 18.1 (0.4) & 24.7 (0.6)& 0.075 (2) & 0.073 (2) &8949.038 (2) &8945.340 (2) & 10.30781 (8) & 0.502 (8) &246.5 (0.2) \\
 CzeV2886            & 77.42 (0.16) &  8031 (fixed) &  7773 (122) & 26.8 (0.7) & 14.6 (0.8) & 58.6 (1.0)& 0.188 (2) & 0.145 (3) &8949.064 (1) &8948.338 (1) &  1.89955 (2) & 0.208 (7) &206.9 (0.5) \\
 CzeV3266            & 85.69 (0.27) & 15400 (fixed) &  6227 (301) & 22.2 (0.4) &  0.3 (0.1) & 77.5 (2.4)& 0.093 (3) & 0.026 (2) &8970.081 (3) &8972.099 (3) &  5.96695 (4) & 0.257 (3) &178.6 (0.8) \\
 CzeV3388            & 84.45 (0.23) &  7197 (fixed) &  5952 (242) & 56.1 (0.8) & 23.3 (1.3) & 20.6 (0.6)& 0.057 (3) & 0.049 (3) &8956.191 (2) &8953.925 (2) & 10.9206 (3)  & 0.479 (4) &  2.7 (0.9) \\
 CzeV3389            & 85.20 (0.71) &  4981 (fixed) &  4070 (170) & 29.1 (0.6) &  5.5 (0.9) & 65.4 (1.3)& 0.163 (7) & 0.111 (5) &8949.774 (1) &8948.602 (1) &  2.95195 (8) & 0.350 (6) &297.3 (0.6) \\
 CzeV3390            & 87.04 (0.35) &  7751 (fixed) &  6426 (313) & 34.1 (0.5) & 13.4 (0.4) & 52.5 (0.7)& 0.075 (2) & 0.062 (2) &8960.325 (1) &8957.055 (1) &  8.4005 (1)  & 0.363 (5) &297.2 (0.3) \\
 OGLE LMC-ECL-06287  & 84.08 (0.16) & 26000 (fixed) & 21340 (267) & 57.7 (1.7) & 39.2 (1.8) &  3.1 (2.9)& 0.050 (8) & 0.049 (8) &4518.851 (4) &4520.258 (4) & 19.0807 (5)  & 0.781 (12)&332.8 (0.4) \\
 OGLE LMC-ECL-06425  & 86.60 (0.22) & 16000 (fixed) & 10318 (412) & 90.2 (1.8) &  9.8 (2.7) &  0        & 0.048 (7) & 0.022 (6) &4520.469 (2) &4519.192 (2) & 16.3146 (4)  & 0.746 (5) &347.0 (0.9) \\
 UCAC4 663-033873    & 84.91 (0.38) &  5869 (fixed) &  5209 (176) & 39.0 (0.9) & 16.7 (0.5) & 44.3 (1.4)& 0.146 (3) & 0.118 (5) &8611.168 (1) &8612.862 (1) &  3.95955 (9) & 0.378 (4) &286.4 (0.6) \\
 OGLE LMC-ECL-09879  & 89.02 (0.09) & 13000 (fixed) & 12490 (230) & 59.6 (0.9) & 40.4 (1.0) & 0         & 0.088 (4) & 0.075 (4) &4513.062 (1) &4506.580 (1) & 13.4545 (3)  & 0.574 (6) &267.7 (0.2) \\
 OGLE LMC-ECL-11563  & 89.00 (0.11) & 18000 (fixed) & 15677 (509) & 54.9 (1.4) & 45.1 (1.3) & 0         & 0.029 (2) & 0.027 (2) &4527.193 (2) &4525.236 (2) & 36.6622 (4)  & 0.802 (5) & 11.2 (0.3) \\
 CzeV3392            & 76.86 (0.29) &  7312 (fixed) &  4803 (412) & 51.1 (0.7) & 11.6 (1.8) & 37.3 (1.3)& 0.206 (4) & 0.199 (6) &8840.653 (1) &8841.174 (1) &  1.4142 (2)  & 0.220 (11)&199.6 (0.9) \\
 CzeV3440            & 86.65 (0.30) & 16500 (fixed) & 11641 (445) & 16.2 (3.0) &  5.4 (0.5) & 78.4 (2.6)& 0.103 (3) & 0.073 (4) &8820.740 (1) &8819.542 (1) &  4.5625 (2)  & 0.459 (5) &323.2 (0.3) \\
 OGLE LMC-ECL-18163  & 89.12 (0.12) & 20000 (fixed) & 14884 (267) & 28.3 (2.4) & 49.6 (4.2) & 22.1 (1.9)& 0.051 (2) & 0.074 (4) &3503.953 (2) &3498.723 (2) & 13.5695 (4)  & 0.698 (6) &280.9 (0.4) \\
 OGLE LMC-ECL-21937  & 86.15 (0.23) & 23000 (fixed) & 20396 (430) & 55.6 (1.1) & 44.4 (3.0) & 0         & 0.022 (1) & 0.021 (1) &2132.178 (2) &2130.102 (2) & 37.1746 (2)  & 0.792 (11)&185.5 (0.2) \\
 OGLE LMC-ECL-24566  & 84.47 (0.20) & 10000 (fixed) &  8713 (178) & 49.6 (0.9) & 50.4 (0.9) & 0         & 0.053 (2) & 0.060 (2) &4512.629 (1) &4514.600 (1) & 29.9319 (2)  & 0.768 (9) &188.0 (0.3) \\
 CzeV3449            & 84.16 (0.11) &  7590 (fixed) &  6475 (102) & 32.1 (1.0) &  8.0 (0.4) & 59.9 (2.3)& 0.156 (3) & 0.098 (4) &8500.640 (1) &8501.604 (1) &  2.48355 (7) & 0.397 (7) &295.0 (0.4) \\
 CzeV3313 = HD 61302 & 89.52 (0.19) &  8200 (fixed) &  7718 (116) & 65.1 (2.3) & 34.9 (1.9) & 0         & 0.038 (2) & 0.031 (2) &7022.190 (3) &7021.173 (3) & 18.9345 (5)  & 0.798 (8) &352.7 (0.6) \\
 CzeV3237            & 89.53 (0.13) &  6328 (fixed) &  6057 (129) & 53.6 (1.1) & 46.4 (1.0) & 0         & 0.044 (1) & 0.044 (1) &4506.619 (1) &4509.957 (1) & 13.8291 (3)  & 0.448 (2) &157.2 (0.2) \\
 TYC 3151-2304-1     & 86.51 (0.09) & 20000 (fixed) & 18570 (244) & 48.0 (0.4) & 43.1 (0.3) & 8.9 (1.1) & 0.066 (2) & 0.066 (3) &8612.836 (1) &8615.202 (1) & 17.1563 (2)  & 0.658 (5) &151.7 (0.1) \\
 CzeV3304            & 76.46 (0.27) &  7045 (fixed) &  4798 (342) & 39.3 (1.6) & 25.8 (1.5) & 34.9 (2.6)& 0.141 (8) & 0.218 (7) &8969.061 (2) &8969.884 (2) & 2.15380 (1)  & 0.201 (9) &202.3 (0.7) \\
 \hline
 \end{tabular}}
\end{table*}

\begin{figure}
  \centering
  \includegraphics[width=0.5\textwidth]{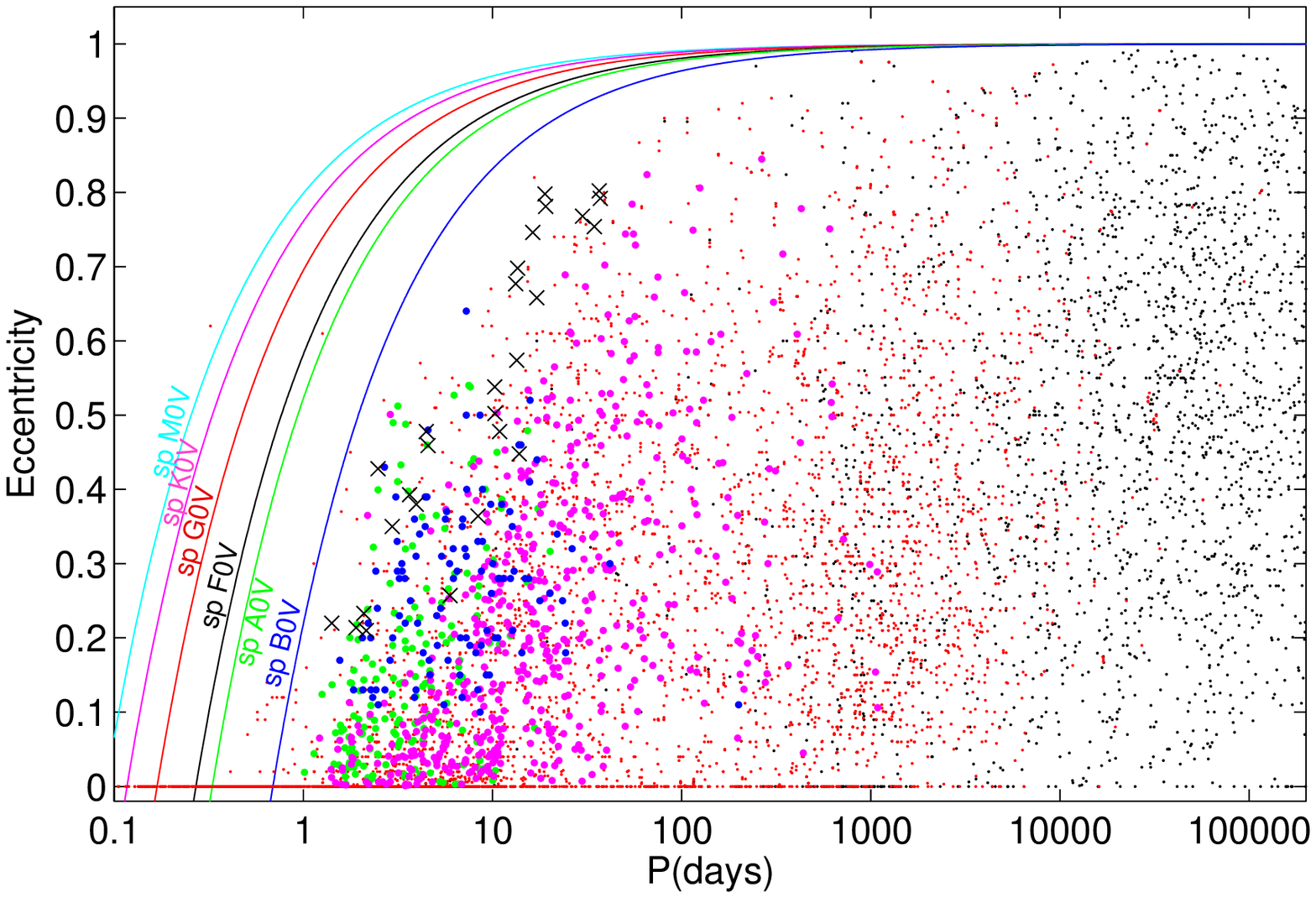}
  \caption{Period-eccentricity diagram. Small black points are the data from the Orbit Catalog of Visual Binaries \citep{2001AJ....122.3472H}, and small red points denote the spectroscopic binaries from the SB9 catalogue \citep{2004A&A...424..727P}. Green dots show eclipsing binaries from the catalogue of eccentric binaries by \cite{2018ApJS..235...41K}, magenta dots show Kepler binaries by \cite{2017AJ....154..105K}, and blue dots show those from ASAS published by \cite{2014MNRAS.441..343S}. The highly eccentric systems of this study are shown by the crosses. The colour curves represent the approximate limits of very close periastron approaches of the two components (i.e. 1.5$\times$ R$_\star$ = $a \cdot (1-e)$ and the semimajor axis $a$ taken from Kepler's third$^{}$ law) when they should collide with each other. The periastron distances were calculated for different spectral types (B to M) according to their typical radii and masses \citep{2013ApJS..208....9P}, with the assumption that the two components are similar to each other (same masses and radii). }
  \label{P-e_diagram}
\end{figure}

\begin{figure}
  \centering
  \includegraphics[width=0.5\textwidth]{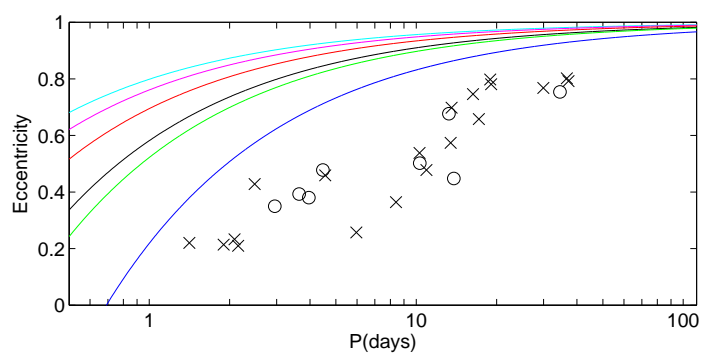}
  \caption{Period-eccentricity diagram of our sample of stars with a distinction between the convective (circles) and radiative (crosses) stars according to the temperature estimates from Table \ref{systemsInfo}. }
  \label{P-e_diagram2}
\end{figure}

\section{Discussion and conclusions}  \label{discussion}

The discovery of several highly eccentric systems is still of great importance for the discussion of
the period-eccentricity diagram (see Fig. \ref{P-e_diagram}) and tidal circularisation theories. The
tendency of closer and more compact systems to orbit on more circular orbits was confirmed here.
Moreover, some of the systems found here lie very close to the upper limit for the eccentricity for a
particular period. These systems should be monitored in more detail in the next years as well.
\cite{1989A&A...223..112Z} and their Fig.1 or \cite{2011MNRAS.411.2804K} and their Fig.4 show that in
serendipitous circumstances, when the eccentricity decrease is fastest, this eccentricity decline
should be of about 0.001/100yr. This lies just at the capability limits of current techniques. After
some time (hundreds or thousands of orbits?), some circularisation should become visible in real time.
This effect should be well observable in precise photometric data, which should later be compared to
our solution. It also should be discussed whether some real change in eccentricity can be detected. The
highly eccentric binaries should be most strongly affected.

The whole situation is clearly more complicated because tidal circularisation processes are differently
effective for early- and late-type stars because of their internal structure (radiative and convective
atmospheres). The two subsets of stars with a distinction between the different temperatures should
therefore show a rather different P-e distribution. However, most of the data from Fig.
\ref{P-e_diagram} do not provide this information and thus cannot be easily divided into these two
different groups of stars. The subset of SBs from \cite{2004A&A...424..727P} might preferentially
contain the hotter stars, and in contrast, most of the EBs from Kepler \citep{2017AJ....154..105K} are
rather later spectral types, but it would be hard to clearly estimate the temperatures for the
eccentric stars from \cite{2018ApJS..235...41K}, or for the long-period visual binaries. Therefore we
plot only our 27 systems in Fig. \ref{P-e_diagram2} in which these two subsets of hotter and cooler
stars differ, and this is still a rather limited sample to show any difference between these two
groups.

On the other hand, when we also take a possible role of the Kozai-Lidov cycles into account, the high
eccentricity value detected here might be a consequence of the triple-star dynamics. The eccentricity
of the inner pair can also be excited by the motion of the third star, and this effect can be much more
rapid than the tidal circularization. Our high relative fraction of the detected systems with third
light would slightly support this hypothesis. Only further investigation of this concern can confirm or
refute any such hypothesis, not only with spectroscopy (convincingly detecting the third component),
but also with good photometry obtained with much better angular resolution to separate the fluxes from
the eclipsing pair itself and some possible close companion. The effect of a close companion that
causes the photocenter shifts can also be studied using the precise TESS data
\citep{2015ApJ...809...77S}, but that is beyond the scope of our analysis. To conclude, our study can
be considered as a first step and calls for more detailed and dedicated observations, especially for
these most eccentric targets in the close future.

\begin{acknowledgements}
At this place we would like to thank the referee of the manuscript, Dr. Andrei Tokovinin, whose helpful
and critical suggestions greatly improved the quality of the manuscript. This research has made use of
the SIMBAD and VIZIER databases, operated at CDS, Strasbourg, France and of NASA Astrophysics Data
System Bibliographic Services. This research made use of Lightkurve, a Python package for TESS data
analysis \citep{2018ascl.soft12013L}.  M.Ma\v{s}ek would like to thank to projects financed by Ministry
of Education of the Czech Republic LM2018102 and LM2018105.
\end{acknowledgements}

\end{document}